\documentclass{emulateapj}

\shorttitle{Blazhko and non-Blazhko RRab stars}
\shortauthors{Jurcsik et al.}

\begin{document}

\title{What is the difference? \\
    Blazhko and non-Blazhko RRab stars and the special case of V123 in M3}

\author{J. Jurcsik\altaffilmark{1}, P. Smitola\altaffilmark{1}, G. Hajdu\altaffilmark{2,3}, C, Pilachowski\altaffilmark{4,*}, K. Kolenberg\altaffilmark{5,6}, \'A. S\'odor\altaffilmark{1,7}, G. F\H {u}r\'esz\altaffilmark{6}, A. Mo\'or\altaffilmark{1}, E. Kun\altaffilmark{8}, A. Saha\altaffilmark{9},   P. Prakash\altaffilmark{10}, P. Blum\altaffilmark{10}, I. T\'oth\altaffilmark{1}}

\affil{$^1$ Konkoly Observatory, H-1525 Budapest PO Box 67, Hungary}
\affil{$^2$ Instituto de Astrof\'{i}sica, Pontificia Universidad Cat\'olica de Chile,
Av. Vicu\~na Mackenna 4860, 782-0436 Macul, Santiago, Chile}
\affil{$^3$ The Milky Way Millennium Nucleus, Av. Vicu\~na Mackenna 4860,
782-0436 Macul, Santiago, Chile}
\affil{$^4$ Department of Astronomy, Indiana University Bloomington, Swain West 319, 727 E. 3rd Street, Bloomington, IN 47405, USA}
\altaffiltext{*}{The WIYN Observatory is a joint facility of the University of Wisconsin-Madison, Indiana University, Yale University, and the National Optical Astronomy Observatory.}
\affil{$^5$ Harvard-Smithsonian Center for Astrophysics. Astronomy, 60
Garden Street MS-42, Cambridge, MA 02138 USA}
\affil{$^6$ Instituut voor Sterrenkunde, K.U. Leuven, Celestijnenlaan 200D, B-3001 Heverlee, Belgium}
\affil{$^7$ Royal Observatory of Belgium, Ringlaan 3, 1180, Brussel, Belgium}
\affil{$^8$ Department of Experimental Physics and Astronomical Observatory, University of Szeged, 6720 Szeged, D\'om t\'er 9, Hungary}
\affil{$^9$ National Optical Astronomy Observatories, Tucson, AZ 85726-6732, USA}
\affil{$^{10}$ Summer interns at the Smithsonian Astrophysical Observatory, Cambridge, MA, 02138, USA}

\email{jurcsik@konkoly.hu}

\begin{abstract}
In an extended photometric campaign of RR Lyrae variables of the globular cluster M3, an aberrant-light-curve, non-Blazhko RRab star, V123, was detected. Based on its brightness, colors and radial velocity curve, V123 is a bona fide member of M3. The light curve of V123 exhibits neither a bump preceding light minimum, nor a hump on the rising branch, and has a longer than normal rise time, with a convex shape. Similar shape characterizes the mean light curves of some large-modulation-amplitude Blazhko stars, but none of the regular RRab variables with similar pulsation periods. This peculiar object thus mimics Blazhko variables without showing any evidence of periodic amplitude and/or phase modulation. We cannot find any fully convincing answer to the peculiar behavior of V123, however, the phenomenon raises again the possibility that rotation and aspect angle might play a role in the explanation of the Blazhko phenomenon, and some source of inhomogeneity acts (magnetic field, chemical inhomogeneity) that deforms the radial pulsation of Blazhko stars during the modulation.
\end{abstract}


\section{Introduction}

RR Lyrae variables are thought to be well-known objects, both from evolutionary and pulsation points of view. They are horizontal-branch stars pulsating in the radial fundamental or in the first overtone mode, and some of them, the double mode RR Lyrae stars, in both. However, the light variation of about 50 per cent of the RRab stars is not stable, they exhibit periodic or complex phase and amplitude modulations \citep[Blazhko effect,][]{bl}.

The light curves of the non-modulated, single-mode RR Lyrae stars are quite regular as the interrelations among their Fourier parameters \citep{jk96} indicate. The applicability of the relations between the physical parameters and light-curve parameters \citep[][and references therein]{j98,kw} verify that the pulsation light curves of the variables reflect their physical properties. Consequently, light curves of RR Lyrae variables with similar physical parameters are similar, and vice versa. Hydrodynamical modeling of the pulsation light curves of RR Lyrae stars is also successful \citep{marconi}.

Some RR Lyrae stars show, however, somewhat anomalous light variations. In some cases, a discrepant evolutionary state is behind the anomalous behavior of these stars. The triple-mode variable AC~And proved to be a large-mass object evolving off the main sequence \citep{acand,kova}. Quite recently, \cite{ogle} found a very low, 0.26 $M_{\Sun}$, `RR Lyr'-like pulsator in a binary system, where mass transfer influenced the stellar evolution of the components. However, it is not at all obvious how to detect the anomalous behavior of the light curve of mono-periodic RR Lyrae stars. In a heterogeneous sample of different age, mass, chemical composition objects, all these parameters influence the shape of the light variation. Therefore, homogeneous groups of objects are ideal targets to find anomalous variables, such as variables in, e.g., globular clusters.

In the present paper, a mono-periodic RR Lyrae star with anomalous light-curve shape in M3 (V123) is displayed and its possible connection with the Blazhko effect is discussed.

\begin{figure*}
\epsscale{1.17}
\plottwo{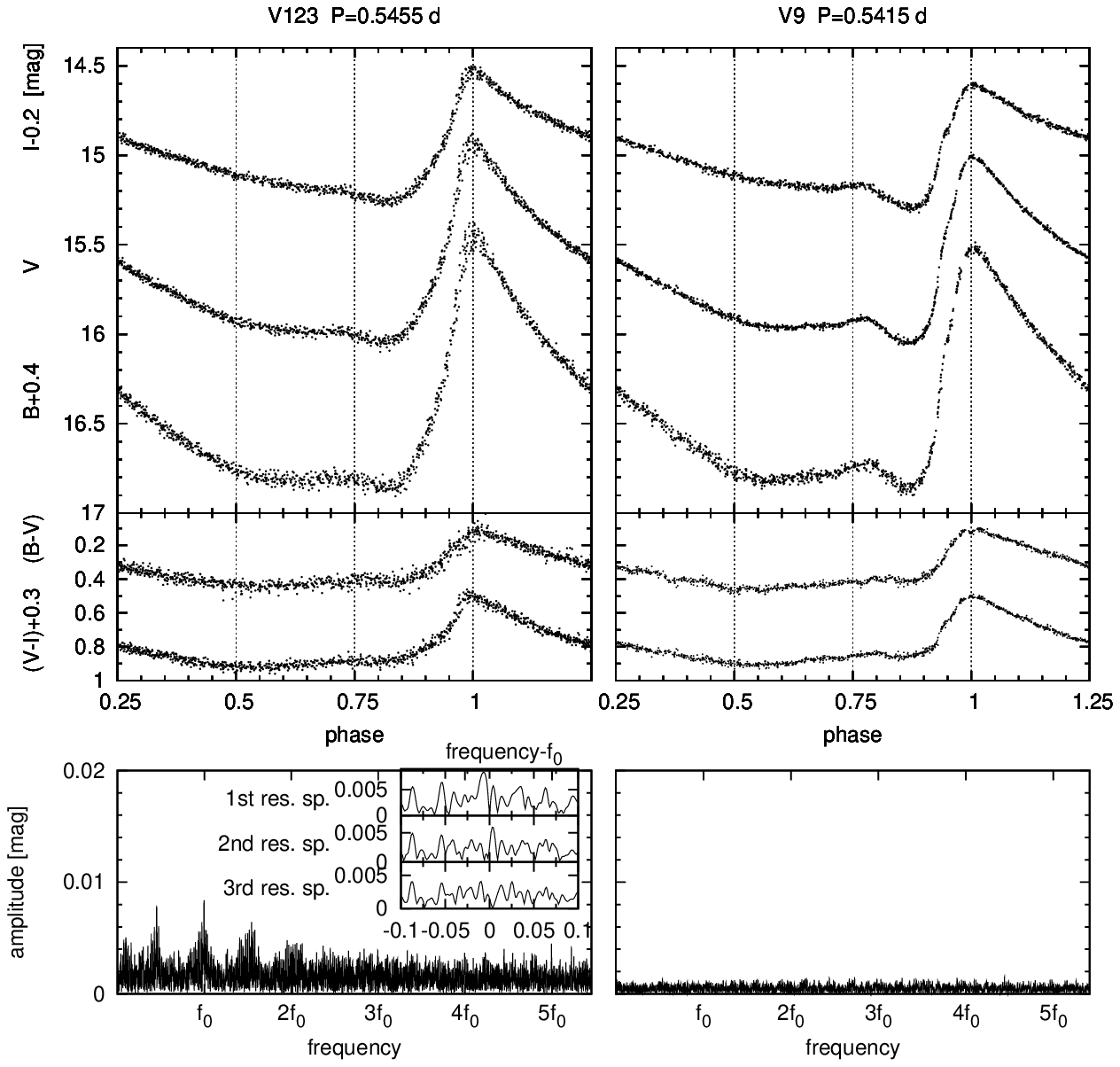}{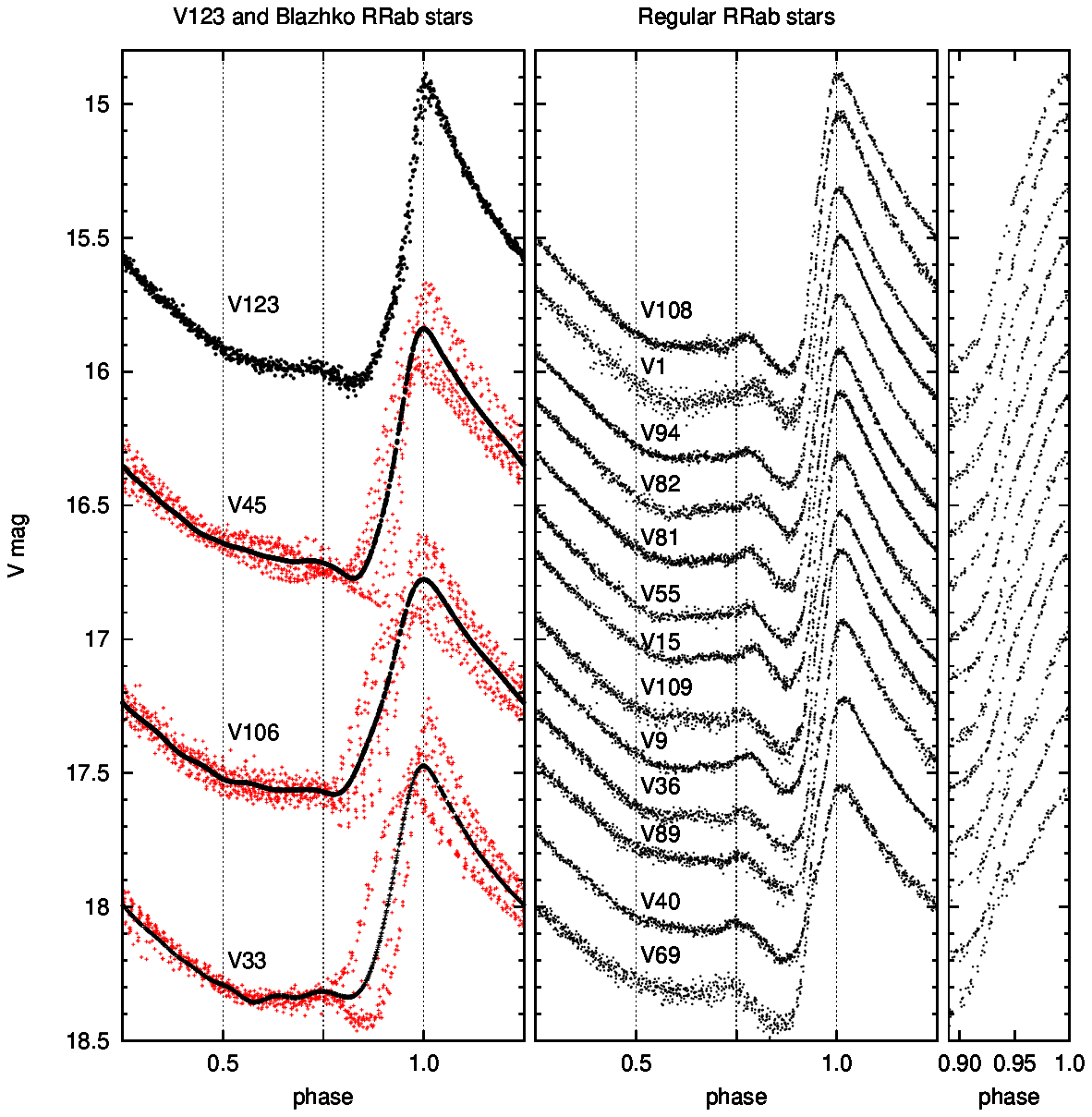}
\caption{Left-side panels: comparison of the  $B, V ,I_{\mathrm C}$ brightnesses and $B-V$ and $V-I_{\mathrm C}$ color curves and the residual spectra after prewhitening for the pulsation frequency and its harmonics of V123 and V9.  The inserts in the bottom-left panel show the vicinities of $f_0$ in consecutive steps of prewhitening for $f_0$ and for the largest amplitude peaks appearing near to $f_0$ in the residual spectra. Right-side panels: comparison of the $V$ light curve of V123 with the light curves of three Blazhko and all the regular RRab stars that have well-measured light curves and periods within the 0.52-0.57 d period range The mean light curves of Blazhko stars are also indicated.  }\label{plotM3}
\end{figure*}
\section{Observations}

One of the most RR Lyrae rich globular clusters, M3, was extensively observed with the 90/60 Schmidt telescope of the Konkoly Observatory in 2012. The data set covers a 200 d long period, and it contains about 1000 measurements in each of the Johnson--Cousins $B,V,I_\mathrm{C}$ photometric bands. Light curves of variables excluding the most crowded regions were obtained from both aperture photometry and image subtraction method techniques.

Spectroscopic observations were obtained using the 
Hydra fiber spectrograph on the 3.5-m WIYN telescope at Kitt Peak  with 
resolving power of $R=3600$ in 2000 \citep{sand}.
A 1000 \AA\ region, centered on the Mg~I triplet lines (5180 \AA)\, was selected 
for observation both 
because of the suitability of the triplet lines for 
radial velocity determinations and because of the relative 
brightness of the RR Lyrae stars at this wavelength. 

Radial velocities for the variables and  red giants were 
measured using the IRAF\footnote{IRAF is distributed by 
the National Optical Astronomy Observatories, which are 
operated by the Association of Universities for Research
in Astronomy, Inc., under cooperative agreement with the National
Science Foundation.} task {\it fxcor}.  
Velocities were determined by cross-correlation relative to the 
twilight sky spectrum observed with the same instrument.
The spectral region for cross-correlation was restricted to
$\lambda\lambda$5000-5400 \AA\ to avoid H$\beta$.

The typical uncertainities of the radial velocity values of variables 
and giants are 12~km~s$^{-1}$ and 1~km~s$^{-1}$, respectively, while 
the uncertanity of the radial velocity zero-point is estimated to 
be less than 1~km~s$^{-1}$ .

The photometric and radial velocity data and full details of the reduction and calibration processes are being given in Jurcsik et al. (in preparation).

\section{V123 in M3}

In the course of the analysis of the data, we have noticed that the light curve of V123 ($V =15.\!\!^{\rm m}70$, $\alpha_{2000} =13^{\rm h}41^{\rm m}52.\!\!^{\rm s}30$, $\delta_{2000} = +28{\degr}06'04.\!\!^{\prime\prime}33$, $P=0.5455$\,d), a non-Blazhko star, is atypical compared to the light curves of similar-period RRab stars. 

The light and color curves of V123 are compared to the light variation of V9 ($P=0.5414$\,d), a typical-light-curve-shape RRab with similar period, in the left panels of Fig~\ref{plotM3}. The differences are evident: in V123, the bump on the lower part of the descending branch is marginal, and there is no hump on the rising branch, while a pronounced bump and a minor hump characterize the light curve of V9. The rising branch of V123 has an anomalous convex shape, it is less steep and the length of the rise time from minimum to maximum is longer than in other RRab stars. 

Although both V123 and V9 are well measured,  separated stars, the scatter of the light curves of the two stars is obviously different. The scatter of the light curves of  V123 arises from minor night to night variations, which can be clearly detected in the variation of the heights of the maximum brightnesses. Nevertheless, the residual spectrum of V123 shown the in the bottom-left panels of Fig~\ref{plotM3}
does not show the typical features of a Blazhko star: the similar equidistant multiplet-frequency structures separated by a well defined modulation frequency at the pulsation frequency, and its harmonics. Instead, although the most prominent residual frequencies appear at $f_0$, their separations are not equidistant, consequently they do not reflect a regular modulation of the light curve. No signal with an amplitude larger than 2 mmag at any half-integer frequency or at the possible positions of the first and second overtone modes is detected.

The $V$ light curves of V123 and three strongly modulated Blazhko stars with similar pulsation periods (V33, P=0.5252 d; V45, P=0.5369 d and V106, P=0.5469 d), and all the regular RRab stars in our sample with periods between 0.52 and 0.57 d  are shown in the right panels of Fig~\ref{plotM3}. The mean light curves of the Blazhko stars, defined by the pulsation components of the Fourier solution including both pulsation and modulation components, are set out. The figure documents clearly that: 1) all the RRab stars plotted in the right panel exhibit very similar light-curve shapes; and 2) the light curve of V123 resembles the mean light curves of Blazhko stars much more than the light curves of normal RRab stars. We also note here that not only the mean light curves of some Blazhko variables look similar to the light curve of V123, but the same is true for the
light curves of some Blazhko stars in a given phase of their modulation.

\begin{figure}
\epsscale{1.0}
\includegraphics{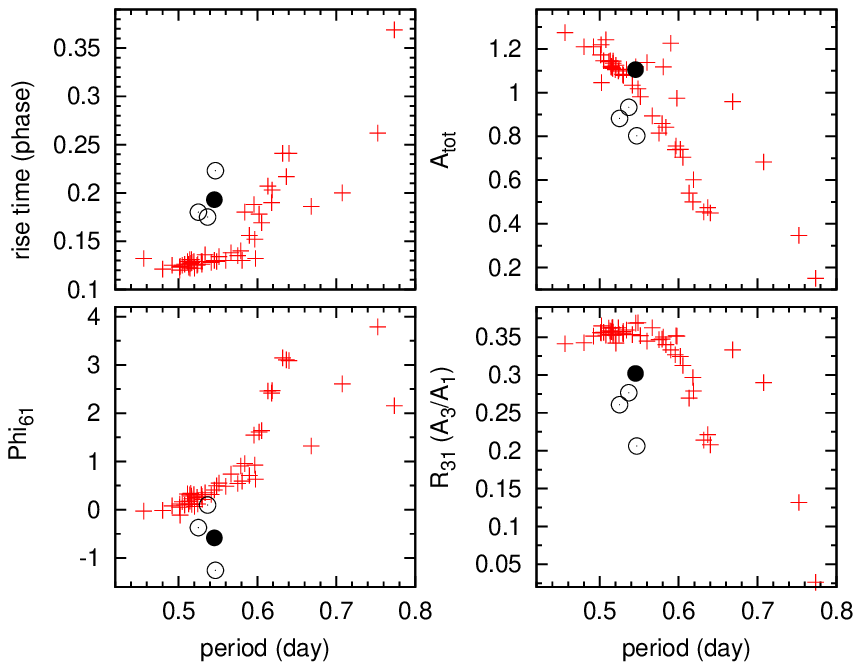}

\includegraphics{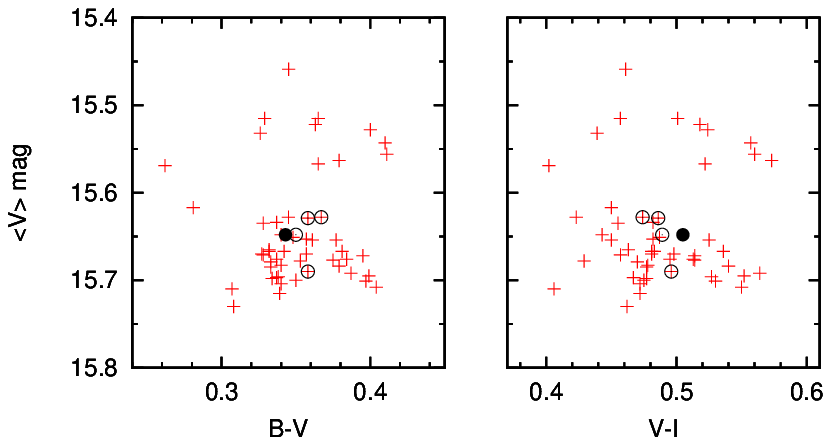}
\caption{Rise time, $A_1, R_{31}$ and $\Phi_{61}$ Fourier parameters of the $V$ light curves of M3 RRab stars are shown in the top and middle panels. Regular variables, V123  and the parameters of the mean light curves of three Blazhko stars shown in Fig~\ref{plotM3} are denoted by '$+$' signal, filled and open circles, respectively.
Intensity-averaged $V$ magnitudes of regular RRab stars in M3 $vs$ magnitude-averaged $B-V$ and $V-I$ colors are plotted in the bottom panels. The positions of V123 and four variables with similar pulsation periods (V9, V36, V4 and V89) are shown by filled and open circles, respectively.}\label{rt-mag}
\end{figure}

We have checked which parameters are responsible for the different shapes of the light curves. It is found  that the rise time and the higher-order amplitude ratios ($R_{k1}=A_k/A_1$, $k>=3$) and phase differences ($\Phi_{k1}=\Phi_{k}-k\Phi_{1}$,$k>=6$) show the most significant discrepancies for V123 and for the mean light curves of the selected Blazhko stars (see top and middle panels in Fig~\ref{rt-mag}). Based on these parameters, V123 clearly stands out from the very homogeneous population of normal RRab stars; instead, it fits the group of the Blazhko stars. Meanwhile, the total amplitudes of V123 and the Blazhko stars are slightly larger and smaller than the total amplitude of normal RRab stars, respectively.

\begin{figure}
\includegraphics[angle=0,scale=.8]{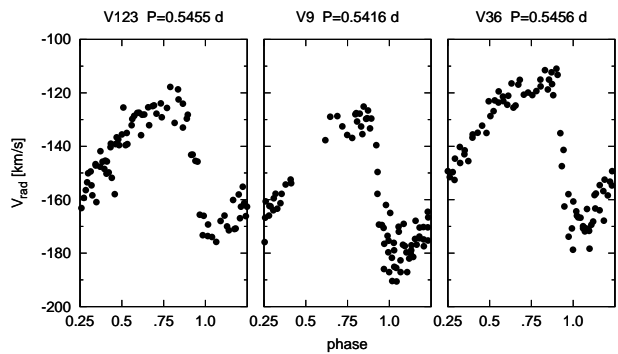}
\caption{Radial-velocity curves of V123, V9 and V36.}\label{rad}
\end{figure}

Another question naturally arises, whether V123 is a bona fide cluster member of M3. The star lies 17 arc sec apart from the cluster's center; it is the largest-radial-distance RR Lyrae star of M3. Its membership is, however, 98 per cent probable, based on the proper motion study by \cite{propm}. The mean magnitudes and colors, and also the radial velocity variation of V123 verify this conclusion as documented in the bottom panels of Fig~\ref{rt-mag} and in Fig~\ref{rad}. The positions of V123 agree within the limits of the uncertainties of the photometry with the positions of not highly evolved, similar-period normal RRab stars in both of the shown color-magnitude diagrams. The  mean radial velocity values of V123, V9 and V36 are also similar, $-146$ km/s, $-154$ km/s and $-138$ km/s, respectively.

The radial-velocity curve of V123 reflects similar anomalies as its light curve, with longer and less steep variation between maximum and minimum as other RRab stars display. The full amplitudes of the radial velocity variations are 50, 55 and 56 km/s for V123, V9 and V36.

Though V123 was not included in the chemical composition analysis of M3 variables \citep{sand} because of uncertain $T_{\mathrm eff}$ and log$g$ information due to the sparseness of the photometric data, a rough analysis of the WIYN spectra revealed that its Fe, Mg and Ca abundances are the same within the errors as the abundances of the other variables.

\begin{figure}
\includegraphics[angle=0,scale=0.8]{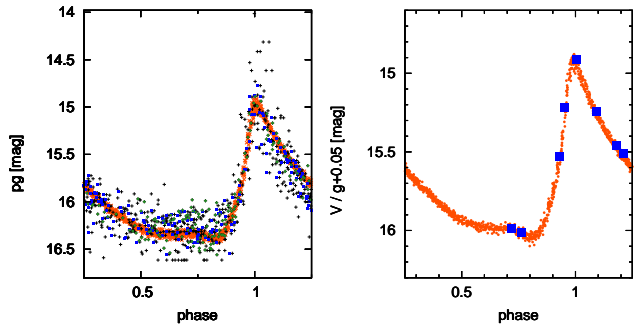}
\caption{Magnitude zero point and time-transformed \citep{oc}  archive photographic (pg) light curves of V123 in comparison with the CCD $B$ data (left panel). The pg and CCD data are shown by `+' and `x' symbols, respectively. Two homogeneous samples of the archive data, the Konkoly observations obtained between 2428900-2437800 and 2438500-2439900 are set out by filled squares and circles. The phase homogenization of the observations takes into account only long-term period changes during the ~100 years of the observations, and does not correct for any possible short-term phase modulation. The  comparison of the 2013 $V$ and the 1999 Thuan-Gunn $g$ \citep{oc} light curves of V123 is shown in the right panel.}\label{123.old}
\end{figure}

Despite the 200 d coverage of the CCD observations, these data do not exclude large amplitude modulation on an even longer time scale. The archive photographic data, however, contradict this possibility. They do not indicate any light-curve variability within the limits of the uncertainty of the data (see Fig~\ref{123.old}). 
The light curve of V123 seemed to be unique with a peculiar-shape rising branch already according to the archive photographic data collected in \cite{oc} without any  sign of a strong light-curve modulation (Blazhko effect). The zero-point and phase homogenized light curves of the different archive observations spanning about 100 years are shown in the left panel of Fig~\ref{123.old}. No  significant variation in the shape of the light curve is evident, the amplitude differences of the data sets arise mostly from the different magnitude scales of the plate materials utilized. Archive CCD observations of V123 are very sparse, only some Thuan-Gunn $uvgri$  data were published in \cite{oc}. The $g$ filter data of V123 match the recent $V$ light curve within the uncertainty limit (right panel in Fig~\ref{123.old}), with a 0.05 mag zero point difference.

We thus conclude that, based on the archive photographic and CCD observations, large-amplitude Blazhko modulation of V123 with a substantially longer modulation cycle than the 200 d interval covered by the recent CCD observations can be certainly ruled out.

\begin{figure}
\includegraphics[angle=0,scale=.9]{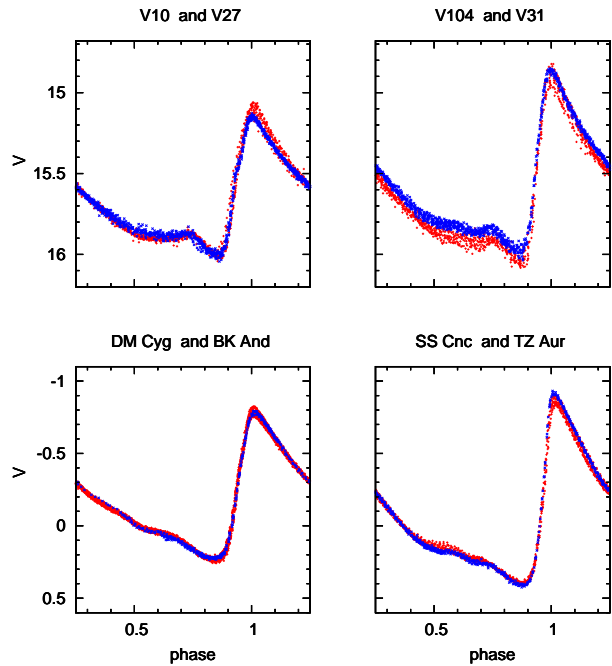}
\caption{Comparison of four pairs of small-modulation-amplitude Blazhko and regular RRab stars with similar periods in M3 (top-left panel: V10, P=0.5696 d and V27, P=0.5791 d; top-right panel: V104, P=0.5699 d and V31, P=0.5807 d) and in the field \citep[bottom-left: DM Cyg, P=0.4216 d and BK And, P=0.4177 d; bottom-right: SS Cnc, P=0.3673 d and TZ Aur, P=0.3917 d. Data of field RRab stars are taken from][]{kbs}. The mean light curves of small-modulation-amplitude Blazhko stars (red/gray symbols) seem to match fairly well the light curves of regular RRab stars (blue/black symbols). The light curves of field stars are set on an arbitrary magnitude scale.}\label{comp}
\end{figure}

V123 is thus either a Blazhko star that exhibits only a small-amplitude, irregular modulation of its anomalous light curve, or it is an anomalous light-curve-shape RRab star showing slight, irregular light-curve fluctuations.

Both the mean and the particular light curves in different phases of the modulation of Blazhko stars are often anomalous. However, the tendency is, that the larger the amplitude of the modulation is, the more  peculiar the light curve can be. Blazhko stars showing small-amplitude modulations look like similar-period RRab stars of regular type as documented in Fig~\ref{comp}. Consequently, it is unlikely that any small-amplitude, irregular modulation accounts for the anomalous shape of the light curve of V123.

Summarizing:
\begin{itemize}
\item[--]{V123 is a cluster member RRab variable of M3 at large radial distance, displaying anomalous-shape light and radial-velocity curves, which resemble the mean variations of Blazhko stars, rather than the variations of similar-period normal RRab stars;}
\item[--]{its mean magnitudes and colors agree well with the magnitudes and colors of RRab stars with similar pulsation periods;}
\item[--]{its light curve shows some irregular variations, but no periodic, strong Blazhko modulation is evident.}
\end{itemize}

\section{Discussion}

What kind of object is V123/M3, this anomalous RRab star?

A possible explanation for the similarity of the light curve of V123 and the mean light curves of Blazhko stars might be the aspect-angle dependence of the shapes of the light curves of Blazhko stars during the modulation cycle. In this case, V123 is, in fact, a Blazhko star seen `pole on',  with a deformed light-curve shape similar to the mean light curve of a Blazhko star, exhibiting slight irregularities only. In \cite{acta}, we have already shown that the Blazhko periods, if converted to rotational velocities, do not contradict the observed $v\sin i$ distribution of horizontal-branch stars. The upper limit for the $v\sin i$ rotational velocities of Blazhko stars defined by spectroscopic studies \cite[$\sim$10km/s,][]{pcl96,cp13} is expected to be exceeded only in the case of extremely short modulation period Blazhko stars ($P<5$ d) with inclination close to 90deg.
These stars are located close to the fundamental blue edge, as short modulation periods are detected only among short pulsation-period variables \citep{acta}.

Such a scenario would mean that the spherical symmetry and/or the homogeneity of Blazhko stars have to be broken. This can be caused by, e.g., magnetic activity, chemical inhomogeneity, or nonradial pulsation modes. 
Although no surface magnetic field of Blazhko stars are detected by recent observations, the deep magnetic field of RR Lyrae stars may remain unobserved according to \cite{stothers}. Observations of the prototype RR Lyr by \cite{chad} showed no evidence for a strong magnetic field in the star's photosphere, and dismissed the hypothesis by \cite{stothers80} that RR Lyr undergoes a magnetic cycle. However, more complex morphologies of a surface magnetic field may remain undetectable with current instrumentation \citep[see ][]{kb}. We may also speculate on that engulfing small companions (cool dwarf stars or planets) on the tip of the giant branch may result surface chemical inhomogeneities in the early stages of the horizontal branch evolution. It is important to note that a detailed high dispersion spectroscopic and photometric study revealed that TY Gru, a large modulation amplitude Blazhko RRab star, shows large over-abundances of carbon and neutron-capture elements, most probably due to mass transfer from an asymptotic giant branch (AGB) companion \citep{p06}. However, no evidence either for a systematic chemical composition difference between Blazhko and non-Blazhko stars or between the observed chemical compositions of Blazhko stars in different phases of the modulation were detected in a detailed spectroscopic study of Blazhko and non-Blazhko stars \citep{fsp11}. Finally, a strong deformation of the main radial pulsation mode by nonradial pulsation components, such as described in the magnetic model by \cite{sh00}, can also contribute to breaking the spherical symmetry of an RR Lyrae star. In a `pole on' configuration, this model results in light variation of  
somewhat anomalous shape without sign of any multiplet components in its Fourier spectrum.

As the light curves of some Blazhko stars during their modulations are temporarily also similar to the light curve of V123, its anomalous behavior might be also interpreted as 
being a Blazhko star, with its pulsation locked in one special Blazhko phase. The explanation of the distorted pulsation curves of Blazhko stars 
during their modulation cycle might be thus the same as for the anomalous light curve shape of V123.

If V123 is neither a `pole-on' nor a `phase-locked' Blazhko star, we have to find the clue why the shape of its light curve is anomalous. As the mean brightness, colors and its Fe, Mg and Ca abundances equal within the errors with the brightness, colors and the abundances of non-evolved regular RRab stars with similar periods in M3, its luminosity and temperature, and as a consequence its mass should not differ significantly from those of other RRab stars. To change the hydrodynamics while maintaining the same main physical properties, we are left again at that the abundances of other chemical compositions of the star  has to be changed significantly to reach any detectable light curve anomaly. This can happen if the evolutionary history of V123 differs from that of other RRab stars. Based on the observed properties, we can exclude, however, the possibility that an evolutionary scenario involving significant mass exchange of a binary system \citep[e.g. in][]{ogle} is behind the phenomenon. If V123 were still a member of a binary system, the secondary should have to be a very low-uminosity object, with no measurable influence on the total luminosity of the system. However, such a companion cannot affect the light-curve shape significantly. If an anomalous chemical composition is behind the aberrant light-curve shape of V123, the only plausible explanation remains the contamination of the atmosphere by the capture of a very small mass object during the evolution on the giant branch. The result of this scenario is, however, most probably an extreme-HB star instead of an RR Lyr \citep{bs}.

Whatever is the solution for V123, its similarity to Blazhko stars suggests that maybe the same mechanism  influences the hydrodynamics and the triggering/damping mechanism of their pulsations, consequently it can be a key object in resolving the Blazhko phenomenon. High-dispersion spectroscopic observations, and hydrodynamical modeling of the anomalous light curve of V123 are needed to reveal the true nature of this peculiar object.

\acknowledgements
This paper is dedicated to the memory of Prof. B\'ela Szeidl, former director of Konkoly Observatory, whose life-long interest on the Blazhko modulation motivated many successful studies.

GH gratefully acknowledges support from the Chilean 
Ministry for the Economy, Development, and Tourism’s Programa Iniciativa Científica Milenio through grant P07-021-F, awarded to The Milky Way Millennium Nucleus. WIYN observations were obtained through the WIYN Queue Program by Paul Smith and Daryl Willmarth, with assistance from G. Rosenstein and W. Hughes;  their contributions are gratefully acknowledged.  CAP acknowledges the generosity of the Kirkwood Research Fund at Indiana University. KK is grateful for the support from a Marie Curie Fellowship (255267 SAS-RRL) within the 7th European Community Framework Programme (FP7). \'AS acknowledges support from the Belgian Federal Science Policy (project M0/33/029).

\end{document}